\documentclass[twocolumn,showpacs,amssymb,nobibnotes,aps,floats,psfig,prb]{revtex4-1}
\usepackage{textcomp,amssymb,graphicx,epsf}
\usepackage{hyperref}
\usepackage{amsbsy}
\usepackage{amsmath}
\usepackage[dvips]{color}
\usepackage{graphicx,epsfig}
\usepackage{epsfig}
\usepackage{float}
\date{}
\def \s{\sigma}
\def \sb{\bar\sigma}
\def \be{\begin{eqnarray}}
\def \ee{\end{eqnarray}}

\def \d{\dagger}
\def \sd{\sum\limits}
\def \sp{\!\!\!}
\def \spb{\sp\sp}
\begin{document}
\title{Correlated singlet phase in the one-dimensional Hubbard-Holstein model}
\author{Sahinur Reja${^1,^2}$}
\author{Sudhakar Yarlagadda${^1}$}
\author{Peter B. Littlewood${^{2,3,4}}$}
\affiliation{${^1}$CAMCS and TCMP Div., Saha Institute of Nuclear Physics,
Kolkata, India}
\affiliation{${^2}$Cavendish Lab, Univ. of Cambridge, Cambridge, UK}
\affiliation{${^3}$Argonne National Laboratory, Argonne IL 60439}
\affiliation{${^4}$University of Chicago,James Franck Institute, Chicago IL 60637}

\date{\today}
\begin{abstract}
We show that a nearest-neighbor singlet phase results
(from an effective Hamiltonian) for the one-dimensional
 Hubbard-Holstein model in the regime of
 strong electron-electron  and electron-phonon 
 interactions and under non-adiabatic conditions
($t/\omega_0 \leq 1$).
By mapping the system  of nearest-neighbor singlets
at a filling $N_p/N$ onto a hard-core-boson (HCB) $t$-$V$
model at 
a filling $N_p/(N-N_p)$,
we demonstrate explicitly that superfluidity and  charge-density-wave (CDW) occur mutually
exclusively with the diagonal long range order manifesting itself only at one-third filling.
{Furthermore, we also 
show that 
the  Bose-Einstein condensate (BEC) 
occupation number $n_0$ for the singlet phase, similar to the $n_0$ for a HCB tight binding model,
 scales as $\sqrt N$; however, the coefficient of $\sqrt N$ in the $n_0$ for the 
interacting singlet phase is numerically demonstrated to be smaller.}
\end{abstract}
\maketitle
\section{Introduction}
The study of coexistence and competition between diagonal long range orders [such as charge density
wave (CDW) and spin density wave (SDW)] 
and off-diagonal long range orders (such as superfluid and superconducting states)
in electronic phases is a subject
 of immense ongoing focus. Of particular interest is the coexistence of
CDW and superconductivity/superfluidity in layered dichalogenides (e.g., 2H-$\rm TaSe_2$,
2H-$\rm TaS_2$, and $\rm NbSe_2$) \cite{withers}, helium-4 \cite{chan},
bismuthates (e.g., $\rm BaBiO_3$ doped with $\rm K$ or $\rm P$) \cite{blanton}, quasi-one-dimensional
trichalcogenide $\rm NbSe_3$ \cite{chaikin} and doped spin ladder cuprate
$\rm Sr_{14}Cu_{24}O_{41}$ \cite{abbamonte},
 quarter-filled organic materials \cite{mori,mckenzie},
non-iron based pnictides (e.g., $\rm SrPt_2As_2$)
{\cite{kudo}, etc. 

Systems with more than one type of interaction typically
manifest a variety of phases of which some cooperate and some compete. A wealth of materials
show evidence of strong electron-phonon (e-ph) interactions besides
 the ubiquitous electron-electron (e-e) interactions. For
instance, transition metal oxides such as cuprates \cite{photoem1,photoem3}
and manganites \cite{lanzara2,
pbl,
boothroyd} and molecular solids such as fullerides \cite{fullarene3}
 indicate strong e-ph coupling. The
interplay of e-e and e-ph interactions in these correlated
systems leads to coexistence of or competition between
various phases such as superconductivity, CDW, SDW, etc.

An archetypal model for understanding the co-occurring effects of e-e and e-ph
interactions is the following well known Hubbard-Holstein model (HHM) \cite{sryspbl1}
\be
H_{hh} \!&=&\! -t\sum_{j\sigma}\left(c^{\dagger}_{j+1\sigma}c_{j\sigma}+ {\rm H.c.}
 \right)
+\omega_0\sum_{j}a_{j}^{\dagger}a_{j}\nonumber\\
\spb && +g\omega_0\sum_{j\sigma}n_{j\sigma}(a_{j}+a_{j}^{\dagger})
+U\sum_{j}n_{j\uparrow}n_{j\downarrow}  ,
\label{ai1}
\ee
where  $c_{j\sigma}^{\dagger}$  is
the fermionic creation operator for itinerant spin-$\sigma$ electrons with hopping integral $t$
and  number operator $n_{j\sigma} \equiv c_{j\sigma}^{\dagger}c_{j\sigma}$,
 $a_{j}^{\dagger}$ is the corresponding bosonic creation operator characterized by a 
dispersionless phonon
frequency $\omega_0$, with $U$ and $g$ representing the 
strengths of onsite e-e and e-ph interactions
respectively.

{ To understand the interplay between the e-e and e-ph interactions, 
the Hubbard-Holstein model
has been extensively studied (in one-, two-, and infinite-dimensions and at various fillings)
by employing various approaches such as
exact diagonalization \cite{exdiag1,exdiag2,exdiag4},
density matrix renormalization group (DMRG)\cite{dmrg,dmrg2},
quantum Monte Carlo (QMC) \cite{qmc1,qmc3,qmc4,qmc5,qmc6,qmc7},
semi-analytical slave
boson approximations \cite{slave_b1,slave_b2,slave_b3,slave_b4,slave_b5},
dynamical mean field theory (DMFT)
\cite{dmft1,dmft2,dmft3,dmft4,dmft5,dmft6,dmft7,dmft8,dmft9},
large-N expansion \cite{largeN2}, variational methods
based on Lang-Firsov transformation \cite{lf1,lf2},   Gutzwiller
approximation
\cite{GA1,GA2}, and cluster approximation \cite{vca}.}

{
In our earlier work\cite{sryspbl1},  in the regimes of strong Coulomb interaction and strong e-ph coupling,
we derived an effective Hamiltonian
 using a controlled analytic approach
that takes into account  dynamical quantum phonons. We solved this effective Hamiltonian
numerically for finite chains and presented a phase diagram for the one-dimensional
Hubbard-Holstein model at quarter filling.
}
It was shown in Ref. \onlinecite{sryspbl1} that while the e-e  interaction produces nearest-neighbor (NN) 
spin antiferromagnetic (AF)
interactions which encourage singlet formation,
 the e-ph interaction generates NN repulsion which is expected to promote CDW order.
It was also shown that a correlated  NN
 singlet phase occurs (at
quarter-filling) and that it carries a signature of a CDW.
In this paper, we  demonstrate that the correlated singlet phase
occurs at other fractions as well and analyze its nature.
Our main result is the demonstration that
 the  NN spin AF and NN repulsive interactions 
compete (instead of cooperate) to produce mutually exclusive (rather than coexisting) 
superfluid 
and CDW phases in the NN singlet phase.
We  show that the  NN singlets 
  manifest superfluidity (and no CDW)
at all fillings that are less than one-half
but not equal to 
one-third and a CDW state (and no superfluidity)
at 
one-third filling.
 Using a modified Lanczos method \cite{sryspbl1,dagotto}
and a newly developed world-line quantum Monte Carlo (WQMC) method we show that the singlet
phase has no Bose-Einstein condensate (BEC) fraction.

In the past, superconductivity due to onsite pairing has been a focus of a number of studies 
\cite{alex,ranninger,hardikar}. Here we are interested in a different situation, namely, NN singlets.
Earlier  a $t$-$J$-$V$ model (involving
bipolarons that are NN singlets)
was introduced 
in Ref. \onlinecite{bonca}.
This $t$-$J$-$V$ model \cite{bonca} [that does
not include the next-nearest-neighbor hopping terms 
   but discusses them qualitatively] is similar to our effective Hamiltonian of Eq. (\ref{Heff})
and can be regarded
as a useful precedent and an endorsement of Eq. (\ref{Heff}).

{ The paper is organised as follows: in Sec. II we briefly derive the effective 
Hamiltonian (that goes beyond the $t-J$ model approximation of the Hubbard model
by including the additional three site residue \cite{eder,troyer,bala}) 
and explain the
various interaction terms and hopping terms. We also point out that
the correlated singlet phase occurs at not only  quarter-filling but also at other fillings.
 In Sec. III, we show
that the correlated singlet phase can be represented by a hard-core-boson (HCB) $t\!\!-\!\!V_1\!\!-\!\!V_2$ model. 
Next, in Sec. IV
we discuss the possibility of formation of a CDW by mapping the $t\!\!-\!\!V_1\!\!-\!\!V_2$ model
onto the well understood $t$-$V$ model. In Sec V, we obtain the 
superfluid density (in the thermodynamic limit) at different filling fractions by using finite size scaling.
In Sec. VI, we analyze the BEC occupation number
at various densities by employing the modified Lanczos method and a newly developed WQMC method.
We close with 
 concluding remarks in Sec. VII.}

\section{Effective HHM Hamiltonian}
We briefly outline below the procedure to get the effective Hubbard-Holstein Hamiltonian
(with more details being provided in Ref. \onlinecite{sryspbl1}). Although we obtain the effective Hamiltonian here in
one-dimension only, our approach is easily extendable to higher dimensions as well.
 We first carry out the
 Lang-Firsov (LF) transformation \cite{LF_transf}
$H^{LF}_{hh}=e^{T}H_{hh}e^{-T}$
where $T=-g\sum_{j\sigma}n_{j\sigma}(a_{j}-a_{j}^{\dagger})$
and get the following LF transformed
Hamiltonian: \\
\be
H^{LF}_{hh}&=& -t\sum_{j\sigma}(X_{j+1}^{\dagger}c_{j+1\sigma}^{\dagger}c_{j\sigma}X_{j}+ {\rm H.c.})+\omega_{0}\sum_{j}
a_{j}^{\dagger}a_{j}\nonumber\\
&& -g^{2}\omega_{0}\sum_{j}n_{j}+(U-2g^2\omega_{0})\sum_{j}n_{j\uparrow}n_{j\downarrow} ,
\label{ai2}
\ee
where $X_{j}=e^{g(a_{j}-a_{j}^{\dagger})}$ and $n_j = n_{j \uparrow} + n_{j \downarrow}$.
Next, we express  as follows our LF transformed Hamiltonian in terms of the composite
fermionic operator
$d_{j\sigma}^{\dagger} \equiv c_{j\sigma}^{\dagger}X_{j}^{\dagger}$:
\be
H^{LF}_{hh}= -t\sum_{j\sigma}\left(d_{j+1\sigma}^{\dagger}d_{j\sigma}+
{\rm H.c.}\right)+\omega_{0}\sum_{j}
a_{j}^{\dagger}a_{j}\nonumber\\
+(U-2g^2\omega_{0})\sum_{j}
n_{j\uparrow}^{d}n_{j\downarrow}^{d}
-g^{2}\omega_{0}\sum_{j}\left(n_{j\uparrow}^{d}+n_{j\downarrow}^{d}\right) ,
\label{ai3}
\ee
where
 $ n_{j\sigma}^{d}= d_{j\sigma}^{\dagger}d_{j\sigma} $.
On dropping the last term, which is a constant polaronic energy, we recognize
that Eq. (\ref{ai3})  essentially represents
the Hubbard Model for composite fermions
 with Hubbard interaction $U_{eff}=(U-2g^2\omega_0)$. In the limit of large $U_{eff}/t$,
using standard treatment involving a canonical transformation,
we get the following effective Hamiltonian written to second order in the small parameter $t/U_{eff}$\cite{eder,troyer,bala}:
\be
H_{t-J-t_3}&=&P_s\left[ -t\sum_{j\sigma}\left(d_{j+1\sigma}^{\dagger}d_{j\sigma}+{\rm H.c.}\right)
+\omega_{0}\sum_{j}a_{j}^{\dagger}a_{j} \right. \nonumber\\
&+&  J\sum_{j}\left(\vec{S}_{j}\cdot\vec{S}_{j+1}-{n_{j}^{d}n_{j+1}^{d}\over{4}}\right)
\nonumber \\
&+&  t_3\sd_{j\s}
\left[d_{j\sb}^{\d}d_{j+1\s}d_{j-1\s}^\d d_{j\sb} + {\rm H.c.} \right]
\nonumber \\
&-& \left . t_3\sd_{j\s}
\left[d_{j\s}^{\d}d_{j+1\s}d_{j-1\sb}^\d d_{j\sb} + {\rm H.c.} \right]\right]P_s ,
\label{ai5}
\ee
where $n_j^{d} =  n_{j\uparrow}^{d}+n_{j\downarrow}^{d}$,
$J={4t^2\over{U-2g^2\omega_{0}}}$, $t_3=J/4$, $\vec{S}_i$ is the spin operator for a spin $1/2$
fermion at site $i$, and $P_{s}$ is the single-occupancy-subspace projection operator.
Furthermore, the last two terms with coefficient $t_3$ ($=J/4$) are the three site terms
which when omitted from the  above Hamiltonian $H_{t-J-t_3}$ yield the well-known $t-J$
Hamiltonian (for the composite
fermionic operators $d_{j\sigma}$).

 The effective $t-J-t_3$ Hamiltonian, given in Eq. (\ref{ai5}), can be re-written in terms of the
original fermionic
 operators $c_{j\sigma}$ as
\be
H_{t-J-t_3}=H_{0}+H_{1} ,
\label{ai7.0}
\ee
where
\be
H_{0}&=& -te^{-g^2}\sum_{j\sigma}P_s \left(c_{j+1\sigma}^{\dagger}c_{j\sigma}+ {\rm H.c.} \right)P_s
+\omega_{0}\sum_{j}a_{j}^{\dagger}a_{j}\nonumber\\
&&+J\sum_{j}P_s\left(\vec{S}_{j}\cdot\vec{S}_{j+1}-{n_{j}n_{j+1}\over{4}}\right)P_s
\nonumber \\
&& +  \frac{Je^{-g^2}}{4}\sd_{j\s}
P_s \left[c_{j\sb}^{\d}c_{j+1\s}c_{j-1\s}^\d c_{j\sb} + {\rm H.c.} \right] P_s
\nonumber \\
&-&  \frac{Je^{-g^2}}{4}\sd_{j\s}
P_s \left[c_{j\s}^{\d}c_{j+1\s}c_{j-1\sb}^\d c_{j\sb} + {\rm H.c.} \right]P_s
\label{ai7.1} ,
\ee
and
\be
\!\!\!\!\!H_{1}&=& -te^{-g^2}\!\sum_{j\sigma}P_s \!\left[c_{j+1\sigma}^{\dagger}c_{j\sigma}(Y_{+}^{j\dagger}
Y_{-}^{j}-1)+ {\rm H.c.}\right]\!P_s .
\label{ai7.2}
\ee
In the above equation, we have separated the $H_{t-J-t_3}$ Hamiltonian into (i) an electronic part $H_{0}$
which is essentially a modified $t-J-t_3$ Hamiltonian containing a NN hopping with a reduced amplitude ($t e^{-g^2}$),
electronic  interaction terms with the same interaction strength $J$,  three site terms with reduced amplitude
$Je^{-g^2}/4$,
and no electron-phonon interaction; and (ii) the remaining perturbative part $H_1$ which corresponds
to the composite fermion terms containing the e-ph interaction with
$Y^{j}_{\pm} \equiv e^{\pm g(a_{j+1}-a_{j})}$. Furthermore, since $J/4 << t$,
we have ignored the following term in $H_1$
\be
&& \!\!\!\! P_s \left [ \frac{Je^{-g^2}}{4}\sd_{j\s}
\left[c_{j\sb}^{\d}c_{j+1\s}c_{j-1\s}^\d c_{j\sb}(Z_{+}^{j\dagger}
Z_{-}^{j}-1) + {\rm H.c.} \right] \right .
\nonumber \\
&& - \left . \frac{Je^{-g^2}}{4}\sd_{j\s}
\left[c_{j\s}^{\d}c_{j+1\s}c_{j-1\sb}^\d c_{j\sb} (Z_{+}^{j\dagger}
Z_{-}^{j}-1)+ {\rm H.c.} \right]\right]P_s ,
\nonumber \\
\label{ai7.1} 
\ee
where $Z^{j}_{\pm} \equiv e^{\pm g(a_{j-1}-a_{j+1})}$.

After carrying out perturbation theory to second-order (as outlined in Ref. \onlinecite{sryspbl1} and Appendix A),
with $t/(g\omega_0 )$ as the small parameter \cite{ravindra},
we get the following effective Hamiltonian:
\be
H_{hh}^{eff}&\cong&-t_{eff}h_{t_1} 
+J h_S -Vh_{nn}
-t_2 h_{\s\s} 
\nonumber \\
&&
-(t_2+J_3) h_{\s \sb} 
+J_3 h_{\s \sb}^{\prime}
\label{Heff}
\ee
{ where 
\be
h_{t_1} = \sd_{j\sigma}P_s \left(c_{j+1\sigma}^{\dagger}c_{j\sigma}+ {\rm H.c.}\right)P_s ,
\label{ht1}
\ee
\be
h_S = \sd_{j}P_s \left
(\vec{S}_{j} \cdot \vec{S}_{j+1}-\frac{1}{4}n_{j}n_{j+1}\right)P_s ,
\label{hS}
\ee
\be
h_{nn} = \sd_{j\s}(1- \!\!n_{j+1\sb})(1- \!n_{j\sb})(n_{j\s}- n_{j+1\s})^2 ,
\label{hnn}
\ee
\be
h_{\s \s} &&= \sd_{j\s}(1-n_{j+1\sb})(1-n_{j\sb})(1-n_{j-1\sb}) \nonumber \\
&& ~~~~~~ \times \left[c_{j+1\s}^\d(1-2n_{j\s})
c_{j-1\s}+ {\rm H.c.} \right] ,
\label{hss}
\ee
\be
h_{\s \sb} 
&& =\sd_{j\s}(1-n_{j+1\sb})(1-n_{j-1\s}) \nonumber \\
&&~~~~~~\times \left[c_{j\s}^{\d}c_{j+1\s}c_{j-1\sb}^\d c_{j\sb} + {\rm H.c.} \right] ,
\label{hss-}
\ee
and
\be
h_{\s \sb}^{\prime} 
&& =\sd_{j\s}(1-n_{j+1\sb})(1-n_{j\s})(1-n_{j-1\sb}) \nonumber \\
&&~~~~~~\times \left[c_{j\sb}^{\d}c_{j+1\s}c_{j-1\s}^\d c_{j\sb} + {\rm H.c.} \right] .
\label{hss-}
\ee
The various coefficients are defined in terms of the system electron-phonon coupling $g$,
the Hubbard interaction $U$, the hopping amplitude $t$, and the phonon frequency $\omega_0$ as follows:  
$V\simeq t^{2}/{2g^2\omega_0}$, $J \equiv {4t^2\over{U-2g^2\omega_{0}}}$, $t_{eff} \equiv t e^{-g^2}$,
$t_2\simeq t^2 e^{-g^2}/{g^2\omega_0}$, and $J_3 = Je^{-g^2}/4$.}
Here the kinetic energy (which is small compared to the interaction energy)
 has contributions from
four hopping terms: $-t_{eff}h_{t_1}$ corresponding to NN hopping 
(with a reduced hopping integral $t_{eff} { \equiv t e^{-g^2}}$),
$-t_2 h_{\s\s}$ representing NNN hopping 
(with double-hopping coefficient $t_2{ \simeq t^2 e^{-g^2}/{g^2\omega_0}}$),  
$-(t_2+J_3) h_{\s \sb}$ implying NN spin-pair $\s\sb$ hopping, and $J_3 h_{\s \sb}^{\prime}$
leading to NN spin-pair $\s\sb$ hopping and flipping to  $\sb\s$; thus  $h_{\s \sb}^{\prime}$
acting on a singlet state produces another singlet state, but with a negative sign.
The NN spin-spin interaction term $J h_S$
{(with $J \equiv {4t^2\over{U-2g^2\omega_{0}}}$)} and
 the NN repulsion term $-Vh_{nn}$ { (with $V\simeq t^{2}/{2g^2\omega_0}$)}
are the dominant
terms in the effective Hamiltonian and compete to form a phase separated cluster at
 larger $J$ (or smaller $U/t$ at a fixed $g$
and $t/\omega_0$). 
 As $J/V$  decreases, the cluster breaks up to undergo a discontinuous transition
to a correlated NN singlet phase as shown in the phase diagram [see Fig. \ref{phasediag}(a)]\cite{adiab0.5}.
 At even lower
values of $J/V$, we get separated single spins (represented by isolated spin phase)
 with the transition at larger $g$ being first-order
while at smaller $g$ it is weakly first order and not continuous
[due to the fact that the system transforms from a superfluid to a CDW, i.e., transition is between
two phases of different symmetry]\cite{sryspbl1}.
{\em The prime objective of the current work is to characterize the correlated singlet state}.

We will now compare the physics related to our effective Hamiltonian, which accounts for various fundamental
processes involved in the kinetic and interaction terms, with the variational Lang-Firsov (LF) treatments 
reported \cite{slave_b4,slave_b5,dmft7,lf2,GA2}.
As the degree of non-adiabaticity decreases, our NNN hopping term $-t_2 h_{\sigma \sigma}$
contribution increases, effectively the hopping transport will  be larger
than that given by $-t_{eff} h_{t_1}$; these two hopping terms together can be regarded as producing a less than
$e^{-g^2}$ suppression of the hopping integral reported in earlier variational LF treatments.
Furthermore, concerning the effect of  including a large Hubbard $U$ term in a Holstein model, we get the 
NN  interaction $2V$  reduced to $2V-J/4$; thus, the mobility would be enhanced which is
consistent again with the earlier works using variational LF transformation.
\begin{figure}[]
\includegraphics[width=0.98\linewidth]{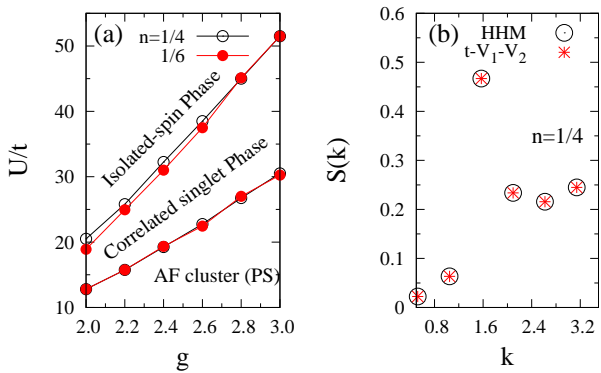}
\caption{(Color online) Plots obtained using modified Lanczos in a twelve-site system for $t/\omega_0 =1 $.
Phase diagram in (a)
 depicts that the phase transition lines are close
for 
both densities $n=1/4$ and $n=1/6$.
Structure factor plots in (b) (drawn at $g=2.2$ and $U/t=17$)
 for the effective 
Hubbard-Holstein model 
(HHM) of Eq. (\ref{Heff})
and the HCB $t$-$V_1$-$V_2$ model  of Eq. (\ref{hcb}) showing that the two models are equivalent.} 
\label{phasediag}
\end{figure}

\section{$t$-$V_1$-$V_2$ hard-core-boson (HCB) model} 
{ In the rest of the paper we study the correlated singlet phase. }
No pair of singlets can share a common site. The closest two singlets can approach
 each other is to have one spin from each singlet be on adjacent sites.
The singlets transport via two processes: (i) the NN spin-pair $\s \sb$  hopping given by the $h_{\s\sb}$
and $h_{\s\sb}^{\prime}$
 terms
in Eq. (\ref{Heff});
and (ii) a second order process involving
breaking of a bound singlet state [with binding energy $E_B = -J+t^2/(g^2 \omega_0)$]
and hopping of the two constituent spins (of the singlet)
 to (a) neighboring sites in the same direction sequentially
[yielding the term $-t_b h_{\s\sb}$ with $t_b \equiv t^2 e^{-2g^2}/|E_B|$]
or (b) neighboring sites in opposite direction and back
[yielding the term $-t_b h_{nn}$].
We now make the important observation that a NN singlet can be represented as a HCB
located at the center of the singlet. Thus the system of NN singlets in a periodic lattice
is transformed into a system of HCB also in a periodic lattice with the same lattice constant $a$
but with the whole lattice displaced by  $a/2$. Then the effective Hamiltonian of the HCB system is
the following $t$-$V_1$-$V_2$ model:
\be
\!\!\!\! \sp\sp\sp H_{b} = \!\sd_{j}  [-T(b^{\dagger}_j b_{j+1} + {\rm H.c.}) + V_1 n_j n_{j+1} +V_2 n_j n_{j+2} ] ,
\label{hcb}
\ee  
where $b_j$ is the HCB destruction operator, $n_j =b^{\dagger}_j b_j$,
$T \equiv(t_2+2J_3+t_b)$, $V_1 = \infty$ (because two singlets cannot share a site), and $V_2 \simeq 2V -J/4$
[with  $V_2 /T > 10$ (i.e., $V_2/T >> 1$) for parameter values in the singlet regime
of our phase diagrams in Fig. \ref{phasediag}(a)]. In the following we set $T=1$.
We corroborate our mapping of the effective 
HHM Hamiltonian
$H_{hh}^{eff}$  (for the singlet phase)
onto the HCB Hamiltonian $H_b$ by demonstrating in Fig. \ref{phasediag}(b) that
the static structure factor
$S(k) \equiv \sum_{l} e^{ikl} W(l)$
for the HHM and HCB cases coincide when the correlation function
$W(l) \equiv (1/N) \sum_{j}[\langle A_j A_{j+l}\rangle -\langle A_j \rangle
\langle A_{j+l} \rangle ]$ is defined through
 $A_j \equiv (S^+_j S^-_{j+1}+ {\rm H.c.})$ for HHM and $A_j \equiv n_j$ for HCB.

It should be made clear that,  for performing calculations,
there is a distinct advantage of accessing bigger system sizes 
for the HCB system as compared to the HHM Hamiltonian. For instance calculations involving
8 HCB (equivalent to 8$\uparrow$ and 8$\downarrow$  electrons)
on a 24 site lattice 
require
{
$\left(
\begin{array}{c}
24\\
8
\end{array}
\right)
 =735471$ basis states in the occupation number representation
and hence
are certainly feasible using modified Lanczos method; on the other hand, using the same technique,
 one can barely deal with 8 electrons (4$\uparrow$ and 4$\downarrow$) on a 16 site lattice for the HHM Hamiltonian
as it requires {
$\left(
\begin{array}{c}
16\\
8
\end{array}
\right)
\times\left(
\begin{array}{c}
8\\
4
\end{array}
\right)
 =900900$} basis states. It is also of interest to note that representing a NN singlet by a HCB located
at the center of the singlet, although has been done here for a one-dimensional system,
can also be done in higher dimensional systems.

\begin{figure}[t]
\includegraphics[width=0.98\linewidth]{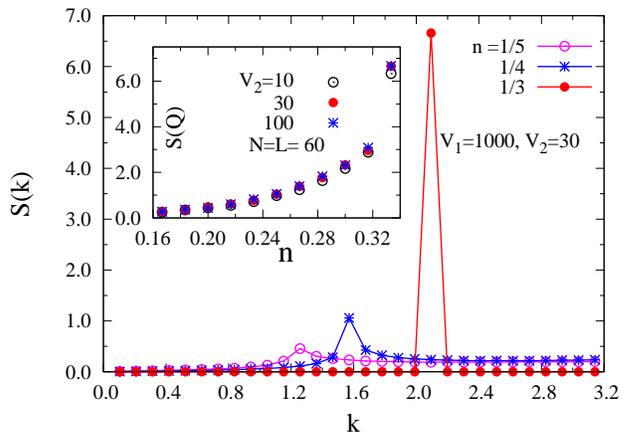}
\caption{(Color online) 
 WQMC plot of the structure factor $S(k)$ versus $k$ -- for $N=L=60$, $\beta = L \Delta \tau$ with 
$\Delta \tau = 0.125$, and at various
densities --  shows CDW at $n=1/3$ with $S(Q) \approx N/9$, i.e., maximum allowed value.
The peak values $S(Q)$ rapidly fall as $n$ moves away from $1/3$ and
are independent of $V_2$ at large values of $V_2$ 
[see inset].}
\label{stf2}
\end{figure}
\begin{figure}[t]
\includegraphics[width=0.98\linewidth]{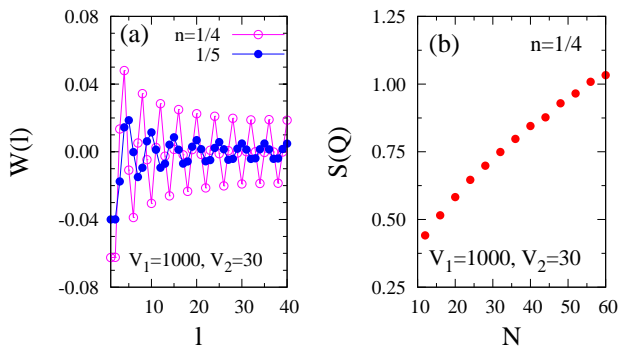} 
\caption{(Color online) Plots, obtained using WQMC at $\beta = N \Delta \tau$
 with $\Delta \tau = 0.125$, showing correlations in the $t$-$V_1$-$V_2$ model.
The correlation function $W(l)$, plotted for $N=80$ sites in (a),
 does not seem to decay. The peak
of the structure factor $S(Q)$, plotted in (b) for various system sizes at $n=1/4$, grows monotonically.}
\label{stf}
\end{figure}

\section{CDW correlations}The repulsive terms in the HCB Hamiltonian $H_b$ indicate that a CDW is possible.
 We study the correlations,
by extending to our $t$-$V_1$-$V_2$ model, 
the well documented WQMC approach for obtaining correlation 
functions and structure factor for the $t$-$V$ model \cite{scalettar1}. 
Plots of the structure factor in Fig. \ref{stf2} show a peak at wavevector $Q = 2\pi n$ suggesting
a CDW.
However (as shown in Fig. \ref{stf2}), only at
filling $n=1/3$, where the structure factor peak is approximately that for the strong CDW case 
corresponding
to $V_2\rightarrow \infty$, can we assert that CDW occurs.
Specifically at $n=1/3$ and for $V_2 > 10$, the $W(l)$ has a simple structure 
[i.e., $W(l) \approx 1/3-1/3\times1/3 = 2/9$ when $l$ is a multiple of 3 whereas for other $l$ values 
$W(l) \approx -1/3\times1/3 = -1/9$]  yielding $S(k) \approx \delta_{k,2\pi/3} N/9 $.
 Furthermore (in Fig. \ref{stf2}), the peak of the structure factor $S(Q)$
 (which remains essentially constant at all relevant
interactions $V_2 > 10$) rapidly decreases as $n$ decreases from $1/3$ -- a trend that is
 similar to that of $S(Q)$
for the $t$-$V$ 
 model
as one moves away from half-filling 
\cite{sdadys}.
Nevertheless, the plots of correlation
function (in Fig. \ref{stf}) do not seem to decay at large distance (for both $n=1/4$ and $n=1/5$)
while the structure factor peak (for $n=1/4$) seems to grow monotonically with system size -- all indicative
 of a CDW.
 Later on, the above ambivalence will be resolved and
it will be demonstrated unequivocally that
our $t$-$V_1$-$V_2$ model has a CDW only at $n=1/3$ while at other fillings $n< 1/3$ superfluidity
(and no CDW) results.

Since $V_1 = \infty$ and because we are dealing with a one-dimensional system, 
we  simplify the phase transition analysis by performing an exact mapping
of the  $N$-site $t$-$V_1$-$V_2$ model onto
a $t$-$V$ model with $N-N_p$ sites and with $V=V_2$. This enables us to access bigger system sizes
for performing numerics; furthermore, since the phase diagram
of the $t$-$V$ model is well known, we can clearly determine the existence of a CDW which
was not possible using the above structure-factor/correlation-function analysis.
Later, we will also show that the $t$-$V$ model lends itself to a
simple finite size scaling approach for obtaining accurately the superfluid density in the thermodynamic limit.

 We first recognize that we can recast the 
HCB Hamiltonian in Eq. (\ref{hcb})
as the following projected Hamiltonian $H_b^P$ where NN sites of a particle are projected out:
\be
\sp\sp\sp H_{b}^{P} = &&\!\sd_{j}  [-T \{(1-n_{j-1})b^{\dagger}_j b_{j+1}(1-n_{j+2}) + {\rm H.c.}\} 
\nonumber \\
&& ~~~~+ V_2 (1-n_{j-1}) n_j (1-n_{j+1}) n_{j+2} (1-n_{j+3}) ]
\nonumber \\
 = &&\!\sd_{j}  [-T (\tilde{b}^{\dagger}_j \tilde{b}_{j+1} + {\rm H.c.}) 
+ V_2 \tilde{n}_j \tilde{n}_{j+2} ].
\label{hcb2}
\ee 
where $\tilde{b}^{\dagger}_j \equiv (1-n_{j-1})b^{\dagger}_j (1-n_{j+1})$ and $\tilde{n}_j \equiv \tilde{b}^{\dagger}_j \tilde{b}_{j}$.
Next, we observe that $H_b^P$ commutes with $\sum_j n_j(1-n_{j+1})$ and thus the total 
number of excitons
(with each exciton comprising of a particle with a hole to its right) is conserved. 
Physically, this is due to the fact that infinite NN repulsion
ensures that the neighboring sites of a particle are unoccupied.
With each particle, we associate only one neighboring vacant site (say, the site on the right side of the particle)
so that situations such as particles on NNN sites can also be dealt with. 
Then by deleting the sites of the holes in all the excitons and having only a NN interaction $V=V_2$
and no other interaction
in the reduced system of $N_1 \equiv N-N_p$ sites, we get the same eigenenergies
(see Ref. \onlinecite{dias} for a similar analysis for the $t$-$V$ model in one-dimension).
 We further
recognize that there is a one-to-one mapping between the eigenstates  of the $ H_b^P$ Hamiltonian 
and the eigenstates of the $t$-$V$ Hamiltonian $H_{t-V}$,
\be
H_{t-V} = \sd_{j}  [-T(b^{\dagger}_j b_{j+1} + {\rm H.c.}) + V n_j n_{j+1} ]  ,
\ee
 with $V =V_2$ and $N_1$
sites while
the corresponding  eigenenergies are identical. We can thus extract 
the eigenenergy spectrum of the $t$-$V_1$-$V_2$  model by studying the equivalent $t$-$V$ model.
We first observe that $n=N_p/N=1/3$ for the $t$-$V_1$-$V_2$ model corresponds to the $n= N_p/(N-N_p)=1/2$
 for the $t$-$V$ model and thus superfluid density vanishes (as the two models have the same eigenenergies)
and
a CDW results \cite{sdadys}
since the mass gap is the same for both. Furthermore, at all fractions $n < 1/3$
for the $t$-$V_1$-$V_2$ model we get a superfluid (and no CDW) since for the $t$-$V$ model
the same is true at $n < 1/2$ \cite{sdadys}. 
Lastly, since $n=1$ for the $t$-$V$ model translates to $n=1/2$ for the $t$-$V_1$-$V_2$ model,
we note that electron-hole symmetry for the $t$-$V$ model guarantees that  $t$-$V_1$-$V_2$ model
exhibits superfluidity and absence of CDW for $1/3 < n < 1/2$ as well.

\begin{figure}[t]
\includegraphics[width=0.98\linewidth]{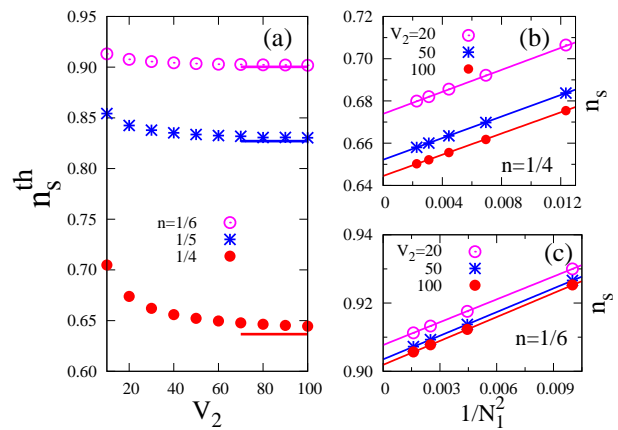}
\caption{(Color online) 
Superfluid density for an infinite system $n_s^{th}$
at various densities $n$ and interactions $V_2$ for the $t$-$V_1$-$V_2$ model
at $V_1 = \infty$ are depicted in (a). 
Values of $n_s^{th}$ in (a) are the intercepts, obtained by extrapolation of the straight lines
through the $n_s$ data plotted at various $1/N_1^2$ values, in figures such as (b) and (c).
{ The solid lines in (a) are for $V_2=\infty$ and obtained 
from Eq. (\ref{V2inf})}.
}
\label{superfluid}
\end{figure}

\section{Superfluid density}
We will now  substantiate
the above observations on the occurrence of superfluidity through calculating the
superfluid density by threading the chain with an infinitesimal magnetic flux. 
We will exploit the one-dimensionality of the system and outline a simple
finite size scaling approach to calculate the superfluid density
in the thermodynamic limit.
 We first note that the
energy for the  $t$-$V_1$-$V_2$ model, when $V_2 = \infty$ and (as before) $V_1 =\infty$, is given by the 
tight binding
Hamiltonian energy for $N_2 \equiv N-2N_p$ particles where we have excluded  
both the NN and NNN holes
 to the right of the particles in the $t$-$V_1$-$V_2$ model. The total energy, 
when threaded by a flux $\theta$,
 is expressed as
\be
E(\theta) = -2T\sd_{k} \cos[k+\theta/N_2] .
\ee
Then the superfluid fraction is given by \cite{fisher,sdsy1}
\be
\!\!\!\!\!\!\!\!
n_s = \frac{N_2^2}{N_p T}\left [ \frac{1}{2} \frac{\partial ^2 E}{\partial \theta^2} \right]_{\theta =0} 
 = \frac{1}{N_p} \frac{\sin \left ( \frac{\pi N_p}{N_2} \right )}{\sin \left ( \frac{\pi}{N_2} \right )} ,
\label{V2inf}
\ee
where anti-periodic (periodic) boundary conditions have been taken for  even (odd) 
values of $N_p$. { The superfluid density in the thermodynamic limit $n_s^{th}$
can be related to the finite ($N_2$-site) system superfluid density $n_s$ as follows:
\be
n_s^{th} = 
 n_s \left [ 1 -
  \frac{1}{6} \left ( \frac{\pi}{N_2} \right )^2 +
  \frac{1}{120} \left ( \frac{\pi}{N_2} \right )^4 ...  \right ] .
\label{ns_th}
\ee
From the above expression (valid for $V_2 = \infty$),
 at a fixed density, 
 we expect $(n_s^{th} -n_s)/n_s \propto 1/N_2^2$ or $1/N_1^2$ (with corrections of 
order $1/N_2^4$ or $1/N_1^4$) for the large but finite
$V_2$ case as well. We calculated the superfluid density at various large values of $V_2$,
system sizes $N$, and filling fractions $n$; we find [as exemplified in Figs. \ref{superfluid}(b) and \ref{superfluid}(c)]
that $n_s$ indeed varies linearly with $1/N_1^2$ using which
we obtain the various $n_s^{th}$ values.}
\begin{figure}[]
\includegraphics[width=0.98\linewidth]{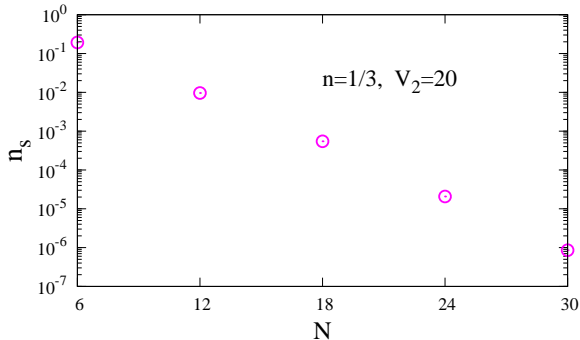}
\caption{(Color online) Superfluid density decaying exponentially with 
system size for the CDW state at
one-third filling and large NNN repulsion $V_2 $.}
\label{exp_sf}
\end{figure} 

 From Fig. \ref{superfluid}(a), we see that the superfluid density
(plotted in the thermodynamic limit)
gradually decreases with increasing $V_2$ and reaches the asymptotic value; 
the $n_s^{th}$ values
for smaller filling fractions
 decrease more slowly
because repulsion is less effective at lower densities.
Regarding the superfluid density at $n=1/3$ and $V_2 = \infty$, it vanishes
at all system sizes as can be seen from Eq. (\ref{V2inf}).
However, at finite $V_2 \ge 10$, $n_s$
 vanishes exponentially with system size [as shown in Fig. (\ref{exp_sf})}] which is
consistent with the fact that
 there is a full CDW gap at $n=1/3$.

\section{BEC occupation number}
{Lastly, we will calculate the Bose-Einstein condensate (BEC) occupation 
number $n_0$. We first recall
the well-established result
that $n_0$, for a system of HCB in a one-dimensional tight binding lattice, varies as $C(n) \sqrt{N}$ in the
thermodynamic limit with the coefficient $C(n)$ monotonically increasing from $0$
as the density $n$ increases
from  $0$ to $1/2$ \cite{lenard,muramatsu1}; consequently, the condensate fraction $n_0/N_p \propto 1/\sqrt{N} \rightarrow 0$.
 Next, in the presence of repulsion (as argued below), we expect  the BEC occupation number $n_0$
to again scale as  $\sqrt{N}$; however, the coefficient of $\sqrt{N}$ will
be smaller due to the restriction on hopping imposed by repulsion.
\begin{figure}[]
\includegraphics[width=0.98\linewidth]{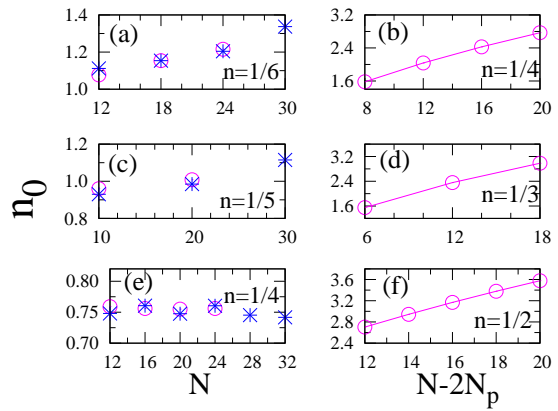}
\caption{(Color online) Plots of BEC occupation number $n_0$, obtained
from modified Lanczos (open circles) and WQMC (crosses), with
(a), (c), and (e) pertaining to $t$-$V_1$-$V_2$ model (with $V_1 = \infty$, and $V_2 = 35$)
while (b), (d), and (f) respectively pertaining to the corresponding 
tight binding  model with enhanced densities $N_p/(N-2N_p)$. For WQMC, $\beta
= N \Delta \tau$ with $\Delta \tau$ = 0.125, 0.15, and 0.175  for (a), (c), and (e)
respectively.}
\label{bec_fig}
\end{figure}

The Bose-Einstein condensate (BEC) occupation 
number $n_0$ 
is obtained from
\be
n_0 = \frac{1}{N}\sd_{i,j} \langle \Psi_0| b^{\dagger}_i b_j|\Psi_0 \rangle ,
\label{bec}
\ee
where $|\Psi_0 \rangle$ is the ground state.
We calculate  $n_0$ using two methods -- modified Lanczos for smaller systems and a newly developed
 WQMC method for both small and larger systems
(see Fig. \ref{bec_fig}). The values of $n_0$ for our $t$-$V_1$-$V_2$
model in a $N$-site original system $S_O$ at various densities [such as $n =1/4,1/5,1/6$] 
seem to be smaller than the $n_0$ for the corresponding transformed tight binding
system $S_{2N_p}$, realized when $V_1 =V_2 = \infty$,
 with $N-2N_p$ sites and enhanced densities
[$n/(1-2n) =1/2, 1/3, 1/4$, respectively].
 This  can be understood from the fact that, in the transformed $S_{2N_p}$ system of
$N-2N_p$ sites  [based on Eq. (\ref{bec})], a particle can hop to 
more sites between two particles than in the original  $t$-$V_1$-$V_2$ system
 leading to a larger $n_0$.  For
the $S_{2N_p}$ system, it is important to realize that $n_0 \propto \sqrt{N-2N_p} \propto \sqrt{N}$.

We will now consider a tight binding system $S_{4N_p}$ with $N-4N_p$ sites and $N_p$ particles
so as to obtain the lower bound for the BEC occupation number $n_0$
for the $N$-site $t-V_1-V_2$ system $S_O$.
For every configuration in the $S_{4N_p}$  system,
there is a corresponding configuration in the 
$S_O$ system that can be obtained by adding two empty sites to the right
and two empty sites to the left of all particles. Furthermore, the ground state  kinetic energy
contribution of the $S_{4N_p}$ and $S_{2N_p}$ systems are both proportional to $N$;
 hence, 
 in the ground state of the original  $S_O$ system, the combined probability weighting of all the configurations
 obtained from the $S_{4N_p}$  system (by adding 4 empty sites next to every particle)
is a finite fraction. Since the BEC occupation  number $n_0$ of $S_{4N_p}$ system 
scales as $\sqrt N$, it follows that the lower bound of the $n_0$ for the
original $S_O$ system also varies as $\sqrt N$. Thus,
the BEC occupation number $n_0$ of the original $N$-site $t-V_1-V_2$ system $S_O$ will
vary as $\sqrt{N}$ since it is constrained from above 
by $n_0 \propto \sqrt N$ for the $S_{2N_p}$ system.

At higher densities (i.e., $1/3 > n \ge 1/5$)
 in  our $t$-$V_1$-$V_2$ model,
we  find that
the values of $n_0$
 seem to increase more slowly with system size 
[see Figs. \ref{bec_fig}(a), \ref{bec_fig}(c), and \ref{bec_fig}(e)] -- 
this being due to smaller coefficients of $\sqrt N$ resulting from interaction effects.
Moreover, we also note [from Figs. \ref{bec_fig}(b) and \ref{bec_fig}(e)] that the 
value of $n_0$ [i.e., the coefficient of $\sqrt N$ in the expression for  $n_0$]
decreases due to repulsion. }

Our new WQMC method (see Appendix B for details) to obtain BEC fraction is a 
modification of the standard approach to 
studying correlations in the xxz model. \cite{scalettar1,scalapino}
Since the Hamiltonian is  real,
it can be shown that the probability amplitude of any basis state in the ground state expression
can be taken as real and non-negative.
Consequently, we approximate the ground state by
\be
 |\Psi\rangle = \sd_i \sqrt {\frac{\langle \phi_i |\exp[-\beta H]|\phi_i \rangle}{Z}} ~ |\phi_i \rangle ,
\ee
with $Z$ being the partition function, $|\phi_i \rangle$ a basis state of the system
in the occupation number representation, and $\beta $ 
being sufficiently large. 
Then we calculate
$n_0$ by setting $|\Psi_0 \rangle = |\Psi\rangle$ in Eq. (\ref{bec}). Our WQMC approach to $n_0$ 
has been benchmarked against the modified Lanczos method
for small system sizes (see Fig. \ref{bec_fig}). The number of passes needed
to estimate $|\Psi\rangle $ turns out to be an order of magnitude larger than that
needed for obtaining correlation functions by WQMC. We take  $|\Psi \rangle$
 to be the state that produces an estimate of the kinetic energy
$\langle \Psi| K |\Psi \rangle$ (with $K$ being the kinetic energy operator) 
that is closest to the usual WQMC estimate $\langle \langle\phi_i| \exp[-\beta H] K|\phi_i \rangle/
\langle\phi_i| \exp[-\beta H]|\phi_i \rangle \rangle_{\rm QMC}$ where $\langle \rangle_{\rm QMC}$
denotes a quantum Monte Carlo average over various states $|\phi_i\rangle$.\\

{
\section{Conclusions}
 In this paper, we have analyzed the correlated NN singlet phase predicted
by the effective Hamiltonian of 
the Hubbard-Holstein model by essentially mapping the Hamiltonian 
onto the well-understood one-dimensional $t$-$V$ model with large repulsion. 
Because the physics is dictated by the $t$-$V$ model,
we find that CDW  and superfluidity occur
mutually exclusively with CDW resulting only
at $n=1/3$ while superfluidity manifests itself at all other fillings. 
{ We also show that the the BEC occupation number $n_0$ for our model scales as $\sqrt N$
similar to the $n_0$ for a HCB tight binding model; additionally, we demonstrate numerically
 (using a new WQMC method and a modified Lanczos
algorithm), at $n \neq 1/3$, that the $n_0$ for our model  
is smaller than the $n_0$
for a HCB tight binding model.} 
}

We close by observing that, while CDW and 
superconductivity seem to be incompatible in the one-dimensional
HHM, experimental results
(such as those reported in Refs. \onlinecite{withers,chan,blanton})
suggest that they can coexist in higher dimensions. 
 Furthermore,
 the vanishing of BEC 
fraction  for the HHM is again an artifact of the one-dimensionality and should
make way to non-zero fractions for higher dimensions just as in the case of the xxz model \cite{sdsy1}. 

\section{Acknowledgments}
S. R. is supported by TCMP \& CAMCS at Saha Institute of Nuclear Physics (India) and
CCT \& COT at Univ. of Cambridge (UK). P. B. L. is supported by the U.S. Department of
Energy under Award No. FWP 70069.

\appendix
\section{}
In this appendix, we will outline our approach to carrying out perturbation theory
and obtaining the ground state energy.
We assume a Hamiltonian of the form $H=H_0+H_1$ where
  the unperturbed $H_0$ has separable eigenstates 
$|n,m\rangle=|n\rangle_{el}\otimes|m\rangle_{ph}$ with  $|0,0\rangle$ being the ground
state with zero phonons; the
 eigenenergies,  corresponding to $|n,m\rangle$, are $E_{n,m}^{(0)}=E_n^{el}+E_{m}^{ph}$. Furthermore,
 the perturbation $H_1$ is the electron-phonon
interaction term of the form given in Eq. (\ref{ai7.2}).\\

   After a canonical transformation\cite{sryspbl1}, we obtain
\be
\tilde H&=&e^{S}He^{-S}\nonumber\\
&=& \!\!H_0\!+\!H_1\!+\![H_0+H_1,S]\!\!+\!\frac{1}{2}\left[[H_0\!+\!H_1,S],S\right] .
\label{ap1}
\ee
In the ground state energy, we know that the first-order perturbation term is zero by construction
(in fact, $\langle n_1, 0|H_1|n_2, 0\rangle =0$).
To eliminate the first-order term in $H_1$, we set $H_1+[H_0,S]=0$. Consequently, we obtain
the matrix elements
\be
\langle n_1,m_1|S|n_2,m_2\rangle=-\frac{\langle n_1,m_1|
H_1|n_2,m_2\rangle}{(E_{n_1,m_1}-E_{n_2,m_2})} .
\label{ap2}
\ee
We now assume that both NN hopping integral $te^{-g^2}$ and
the Heisenberg spin interaction strength $J$ are much smaller
compared to the phononic energy $\omega_{0}$ which is true
at large couplings $g$. Hence, we make the approximation
$(E_{n_1,m_1}^{(0)}-E_{n_2,m_2}^{(0)}) \simeq (E_{m_1}^{ph}-E_{m_2}^{ph})$;
then, using Eqs. (\ref{ap1}) and (\ref{ap2}), we obtain
\begin{widetext}
\be
{_{ph}\!\langle m_1|\tilde H|m_2\rangle_{ph}}  
\simeq
 {_{ph}\!\langle m_1|H_0|m_2\rangle_{ph}}+\frac{1}{2}\sum_{\bar m}
{{_{ph}\!\langle m_1|H_1|\bar m\rangle_{ph}}~
{_{ph}\!\langle \bar m|H_1|m_2\rangle_{ph}}}
\left [ \frac{1}{E_{m_2}^{ph}-E_{\bar m}^{ph}}
+\frac{1}{E_{m_1}^{ph}-E_{\bar m}^{ph}} \right ]
.
\label{ap3}
\ee
Next, it is important to note that the second order correction $E^{(2)}_{n,m}$,
corresponding to the unperturbed eigenenergy $E_{n,m}^{(0)}$,
can be expressed as follows:
\be
E^{(2)}_{n,m} =
\sum_{ \bar m}
\frac{{\langle n, m|H_1| \bar m\rangle_{ph}}~
{_{ph}\!\langle \bar m|H_1|n, m\rangle}}
{{E_{m}^{ph}-E_{\bar m}^{ph}}}
\simeq {\langle n, m|\tilde H|n, m\rangle} -  {\langle n, m|H_0|n, m\rangle}
.
\label{ap4}
\ee
\end{widetext}
Furthermore, since $\langle n_1, 0|H_1|n_2, 0\rangle =0$, $\langle n, 0|\tilde H|n, 0\rangle$ is the total energy 
that resulted from performing second order perturbation theory on the unperturbed energy $E_{n,0}^{(0)}$.
 Our procedure for finding ground state
amounts to obtaining the lowest eigenvalue for the matrix with elements $\langle n_1, 0|\tilde H|n_2, 0\rangle$; this
is equivalent to finding the ground state of the effective Hamiltonian $H_e$ (as was done in 
 Ref. \onlinecite{sryspbl1}):
\be
H_e ={_{ph}\!\langle 0|H_0|0\rangle_{ph}}+ H^{(2)} ,
\ee
where
\be
H^{(2)} = \sum_{ \bar m}
\frac{{_{ph}\langle 0|H_1| \bar m\rangle_{ph}}
\times{_{ph}\!\langle \bar m|H_1|0\rangle_{ph}}}
{{E_{0}^{ph}-E_{\bar m}^{ph}}} .
\ee
 This procedure amounts to considering the restricted subspace
spanned by eigenstates 
 $|n,0\rangle_{1}$
obtained from carrying out first order perturbation theory on $|n,0\rangle$:
\be
|n,0\rangle_{1} = |n,0\rangle +
\sum_{ \bar m} 
\frac{{| \bar m\rangle_{ph}}
~{_{ph}\!\langle \bar m|H_1|n, 0\rangle}}
{{E_{0}^{ph}-E_{\bar m}^{ph}}}
,
\label{ap5}
\ee
It is important to recognize that the state $|n,0\rangle_{1}$ is
 not separable, i.e., cannot be expressed as a product of an electronic wavefunction and a phononic wavefunction.
We have restricted ourselves to the subspace of the states $|n,0\rangle_{1}$ because the states
$|n,m \neq 0\rangle_{1}$ correspond to higher energy states due to the fact that the electronic
excitation energy is much smaller than the phononic energy, i.e., $t e^{-g^2} << \omega_0$.
Additionally, we would like to point out that the total ground state energy (in second order perturbation theory)
is obtained by diagonalizing the matrix whose elements are $\langle n_1, 0| H|n_2, 0\rangle_{1}$.

{{
\section{WQMC FOR BEC FRACTION}
We will discuss, in brief, the usual world-line quantum Monte Carlo (WQMC)
approach  \cite{scalettar1,scalapino}
adapted for calculating correlations
in our $t$-$V_1$-$V_2$ model Hamiltonian given below:
\be
H_{b} = \!\sd_{j}
 { H^j }&=& \sd_j [-T(b^{\dagger}_j b_{j+1} + {\rm H.c.}) \nonumber\\
&+&V_1 n_j n_{j+1} +V_2 n_j n_{j+2} ] .
\ee
Since this is quite similar to the $t$-$V$ model, we can employ the checkerboard decomposition 
$H_b=H_1+H_2$ where
{$H_1=\!\sd_{j~{\rm odd}}  H^j $
and
$H_2=\!\sd_{j~{\rm even}}  H^j $.}
It is important to note that both $H_1$ and $H_2$ consist of independent two-site pieces.
Because of the decomposition, it becomes easier to evaluate the expectation value of an operator $A$
given by
\be
\langle A\rangle=\frac{Tr[Ae^{-\beta H_b}]}{Tr[e^{-\beta H_b}]} ,
\ee
with $A$ involving only number operators (such as $n_i n_j$) or NN hopping
operators (such as $b^{\dagger}_j b_{j+1}+ {\rm H.c.}$).
 Now we calculate the partition function:\\
\be 
Z &=& Tr[e^{-\beta H_b}]
 \nonumber \\
&=&\sd_{i_1,...,i_{2L}}\!\!\!\!\langle i_1|U_1|i_{2L}\rangle\langle i_{2L}|U_2|i_{2L-1}\rangle...
{\langle i_3|U_1|i_{2}\rangle}
\langle i_2|U_2|i_{1}\rangle. \nonumber
\ee
Here $U_{i}=e^{-\Delta \tau H_i}, \beta=L\Delta\tau$, and each of
$|i_1\rangle$, ...,$|i_{2L} \rangle$ 
form a complete 
basis set in the occupation number representation.
Here the world lines  are the locus of the particles in the imaginary 
time ($\tau$) direction.

For the density-density correlation function
$\langle n_i n_{i+l}\rangle$ (which is the expectation value of a diagonal operator), 
the above procedure of inserting $2L$ time slices yields the simple form
\be
\langle n_in_{i+l}\rangle=\frac{1}{2}\left\langle[\langle i_L|n_in_{i+l}|i_L\rangle +
\langle i_{L+1}|n_in_{i+l}|i_{L+1}\rangle]\right\rangle_{\rm QMC}\nonumber   ,
\ee
where $\langle\hspace{3mm} \rangle_{QMC}$ represents average over many QMC passes.
Notice that we have concentrated only on $L$ and $L+1$ time slice indexes although
expectation value can be taken over all the $2L$ time slice indexes for better statistics.
As for $\langle b^{\dagger}_j b_{j+1}+ {\rm H.c.}\rangle$ (which corresponds to a non-diagonal operator),
WQMC procedure yields
{
\be
\langle b^{\dagger}_j b_{j+1}+ {\rm H.c.}\rangle=
\langle \frac{
\langle i_M|(b^{\dagger}_j b_{j+1} + {\rm H.c.})U_k|i_{M+1}\rangle} 
{\langle i_M|U_k|i_{M+1}\rangle}\rangle_{\rm QMC}\nonumber ,
\ee
where,
 for odd (even) values of $j$, we take $k=1$ ($2$) and even (odd) $M$.}
However, as regards obtaining expectation value of $( b^{\dagger}_j b_{j+m}+ {\rm H.c.})$ for $m > 1$,
the simple procedure (involving checkerboard decomposition) given above is not applicable;
moreover, other suggested procedures in the literature are complicated \cite{scalettar1}.

Here, we propose an alternate simple method for evaluating 
$ \langle b^{\dagger}_j b_{j+m}+ {\rm H.c.} \rangle$ for $m > 1$
and thus obtaining the BEC occupation number
\be
n_0 = \frac{1}{N}\sd_{i,j} \langle \Psi_0| b^{\dagger}_i b_j|\Psi_0 \rangle ,
\label{bec_app}
\ee
with $|\Psi_0 \rangle$ being the ground state.
To the WQMC method mentioned above,
 we add our trick to construct $|\Psi_0 \rangle$
as a linear combination
of the basis states $|\phi_i \rangle $
in the occupation number representation, i.e.,
 $|\Psi_0\rangle = \sd_i a_i|\phi_i \rangle$ with $\sd_i a_i^2=1$.
Once we get a good estimate of the ground state $|\Psi_0\rangle$, we can
calculate the expectation values of any operator.

After equilibrium (which is attained after several QMC passes),
 we run the simulation for a sufficient number of QMC passes 
and store  the
 basis states  corresponding to time slices $L$ and $L+1$ in each pass. 
It is obvious that 
some of the basis states will
occur more frequently.
The frequency of occurrence of a basis state $|\phi_i \rangle $
is proportional to the probability ($a_i^2$) of
its occurrence in the expansion of the ground state $|\Psi_0 \rangle$.
Now, the coefficients $a_i$ can be taken
as real because the Hamiltonian is real and 
consequently $|\Psi_0\rangle$ can also be taken as real.
Furthermore, all $a_i$ can be taken to be positive for the following reason.
Firstly, the expectation values of NN and NNN interaction terms remain unaffected by the sign of $a_i$ . 
Next, the
expectation value of the hopping term is given by
\be
-T\langle \Psi_0|b_l^{\dagger}b_{l+1}|\Psi_0\rangle&=&\nonumber
-T\sd_{i,j}\langle \phi_i|a_i (b_l^{\dagger}b_{l+1})a_j|\phi_j \rangle]\nonumber\\
&=&-T\sd_{i,k}\langle \phi_i|a_i c_k|\phi_k \rangle]\nonumber\\
&=&-T\sd_{i}a_i c_i   .
\ee
This value is minimized when $a_i$ and $c_i$ have the same sign. Then, if we take
$a_i$ to be positive for all $i$, $c_i>0$ for all $i$. Thus in 
$|\Psi_0\rangle = \sd_i a_i|\phi_i \rangle$, we can take all $a_i$
to be positive and real.

 Let $|\Psi_i \rangle $ and $E_i$ be the eigenstates 
and the eigenenergies of the Hamiltonian
with  $E_0$ being the ground state energy.
 For sufficiently large $\beta$, we approximate the ground state by 
\be
 |\Psi\rangle &=& \sd_i \sqrt {\frac{\langle \phi_i |\exp[-\beta H]|\phi_i \rangle}{Z}} ~ |\phi_i \rangle ,
\ee
because then
\be
|\Psi\rangle &=& \sd_i \sqrt {\frac{\langle \phi_i |\sum_j |\Psi_j \rangle \langle \Psi_j |\exp[-\beta H]
\sum_k |\Psi_k \rangle \langle \Psi_k ||\phi_i \rangle}{Z}} ~ |\phi_i \rangle
\nonumber \\
 &\approx & \sd_i \sqrt {\frac{\langle \phi_i |\Psi_0 \rangle \exp[-\beta E_0]
 \langle \Psi_0 |\phi_i \rangle}{Z}} ~ |\phi_i \rangle
\nonumber \\
&\approx & \sum_i  \langle \phi_i |\Psi_0 \rangle |\phi_i \rangle
= |\Psi_0 \rangle
,
\ee
since the partition function
$Z= \sum_i \langle \Psi_i |\exp[-\beta H]|\Psi_i \rangle \approx \exp[-\beta E_0]$.
}}


\begin{thebibliography}{1}
\bibitem{adiab0.5}
The phase diagram results for $t/\omega_0 =0.5$ are qualitatively
similar to those for $t/\omega_0=1.0$.
\end{thebibliography}
\end{document}